\documentclass[10pt,twocolumn,aps,prl,floatfix,superscriptaddress,showpacs]{revtex4-1}
\usepackage{amsmath,amsfonts,amssymb}
\usepackage{bm}
\usepackage{epsfig}
\usepackage{float}
\usepackage{graphicx}
\usepackage{hyperref}
\usepackage{natbib}

\def\et{\epsilon_T}
\def\mt{\mu_T}
\def\ea{\epsilon_A}
\def\eac{\ea^*}
\def\eat{\bar{\epsilon}_A}
\def\ma{\mu_A}
\def\pn{p_{\text{new}}}
\def\pa{\rho_A}

\begin{document}

\title{Opinions, Conflicts and Consensus: Modeling Social Dynamics in
a Collaborative Environment}
\author{J\'anos T\"or\"ok}
\affiliation{Institute of Physics, Budapest University of Technology
and Economics, H-1111 Budapest, Hungary}
\author{Gerardo I\~niguez}
\affiliation{Department of Biomedical Engineering and Computational
Science, FI-00076 Aalto, Finland}
\author{Taha Yasseri}
\affiliation{Institute of Physics, Budapest University of Technology
and Economics, H-1111 Budapest, Hungary}
\author{Maxi San Miguel}
\affiliation{IFISC (CSIC-UIB), Campus Universitat Illes Balears,
E-07071 Palma de Mallorca, Spain}
\author{Kimmo Kaski}
\affiliation{Department of Biomedical Engineering and Computational
Science, FI-00076 Aalto, Finland}
\author{J\'anos Kert\'esz}
\affiliation{Center for Network Science, Central European University,
H-1051 Budapest, Hungary}
\affiliation{Institute of Physics, Budapest University of Technology
and Economics, H-1111 Budapest, Hungary}
\affiliation{Department of Biomedical Engineering and Computational
Science, FI-00076 Aalto, Finland}

\date{\textrm{\today}}

\begin{abstract}
Information-communication technology promotes collaborative
environments like Wikipedia where, however, controversiality and
conflicts can appear. To describe the rise, persistence, and
resolution of such conflicts we devise an extended opinion dynamics
model where agents with different opinions perform a single task to
make a consensual product. As a function of the convergence parameter
describing the influence of the product on the agents, the model shows
spontaneous symmetry breaking of the final consensus opinion represented
by the medium. In 
the case when agents are replaced with new ones at a
certain rate, a transition from mainly consensus to a perpetual
conflict occurs, which is in qualitative agreement with the scenarios
observed in Wikipedia.
\end{abstract}

\maketitle


Society represents a paradigmatic example of complex systems, where
interactions between many constituents and feedback and other
non-linear mechanisms result in emergent collective phenomena. The
recent availability of large data sets due to
information-communication technology (ICT) has enabled us to apply
more quantitative methods in social sciences than before. As a matter
of fact, physicists play an increasing role in the study of social
phenomena by applying physics concepts and tools to investigate
them~\cite{castellano2009statistical}.

Social interactions are heavily influenced by the opinions of the
members of the society. This is especially true when complex tasks
are to be solved by cooperation, as practiced throughout the history
of mankind.  In this respect new technologies open up unprecedented
opportunities: by using the Internet and related facilities even
remote members of large groups can work on the same task and achieve a
higher level of synergy. Examples include open software projects or
large collaborative scientific endeavors like high-energy physics
experiments. However, it is unavoidable that in such cases differences
in attitudes, approaches, and emphases (in short, opinions) occur.
Then questions arise: How can a task be solved in a collaborative
environment of agents having diverse opinions? How do conflicts emerge
and get resolved? The understanding of these mechanisms may lead to an
increase in efficiency of value production in cooperative
environments.  A prime example of the latter is Wikipedia, a free,
web-based encyclopedia project where volunteering individuals
collaboratively write and edit articles on their desired topics.
Wikipedia is particularly well-suited also as a target for a wide range of
studies, since all changes and discussions are recorded and made publicly
available~\cite{zlatic2006wikipedias,capocci2006preferential,wilkinson2007assessing,kittur2007says,ratkiewicz2010characterizing,yasseri2012circadian}.

\begin{figure}[t]
\centering
\includegraphics[width=\columnwidth]{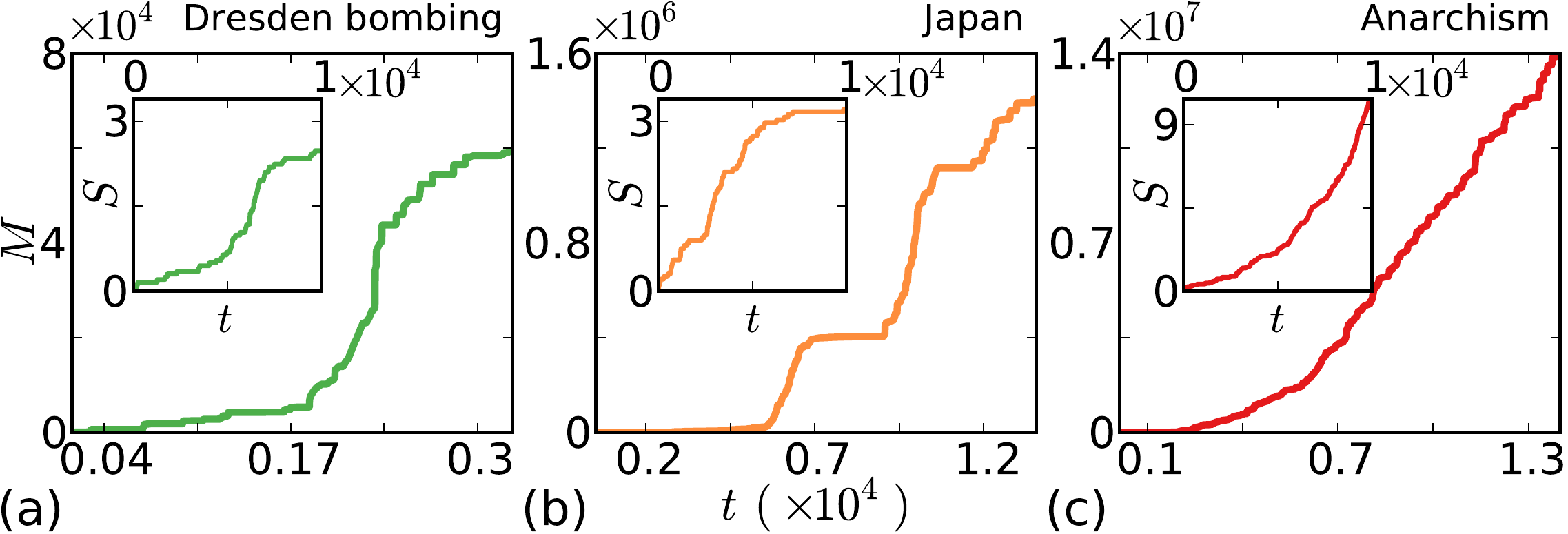}
\caption{(Color online) 
Empirical controversiality measure $M$~\cite{yasseri2012dynamics} as a function of the
number of edits $t$ for three different conflict scenarios in Wikipedia, corresponding
to: (a) single conflict, (b) plateaus of consensus, and (c) uninterrupted controversiality.
Titles are the article topics. Inset: Theoretical conflict measure $S(t)$ of Eq.~(\ref{eq:sumChanges}),
allowing for a qualitative comparison of the model with the empirical results.}
\label{fig:empirical}
\end{figure}

Recently, the controversiality and dynamical evolution of Wikipedia
articles have been studied in detail and typical patterns of different
categories of the so-called {\em edit wars} have been
identified~\cite{sumi2011characterization,sumi2011privacy,yasseri2012dynamics,vuong2008ranking,brandes2008visual}.
Take for example Fig.~\ref{fig:empirical}, where the evolution of a controversiality
measure $M$ based on mutual reverts and maturity of editors is shown for three
different regimes of conflict.

Our aim in this Letter is to model controversiality and conflict resolution
in a collaborative environment and, where possible, to qualitatively
compare our results with different scenarios observed in the Wikipedia.
One of the most developed areas of quantitative modeling in social phenomena is opinion
dynamics~\cite{castellano2009statistical,holyst2001social,haoxiang2011opinion}.
These models have much in common with those of statistical physics, yet
interactions are socially motivated. The problem with these models is,
similarly to evolutionary game theoretic ones~\cite{szabo2007evolutionary},
that results are usually evaluated by qualitative subjective judgment
instead of comparison with empirical data, one exception being the study of
elections~\cite{bernardes2002election,fortunato2007scaling}. The well-documented
Wikipedia edit wars can also be of use in this respect. 


Our model is based on the \textit{bounded confidence} (BC)
mechanism~\cite{deffuant2000mixing,hegselmann2002opinion},
describing a generic case when convergence occurs only upon opinion
difference being smaller than a given threshold. Here $N$ agents are
characterized by continuous opinion variables $x_i{\in}[0, 1]$ and
their interactions are pairwise. The agents' mutual influence is
controlled by the convergence parameter $\mt$, only if their
opinions differ less than a given tolerance $\et$, i.e. for
$|x_i{-}x_j|{<}\et$ we update as follows,
\begin{equation}
\label{eq:deffuant}
(x_i, x_j) \mapsto (x_i + \mt[x_j - x_i], x_j + \mt[x_i - x_j]).
\end{equation}

The dynamics set by Eq.~(\ref{eq:deffuant}) has been studied
extensively in the literature~\cite{castellano2009statistical},
initially by using the mean-field approach of two-body inelastic
collisions in granular
gases~\cite{ben2000multiscaling,baldassarri2002influence}. It leads to
a frozen steady state which is characterized by $n_c{\sim}1/(2\et)$
disjoint opinion groups. This is caused by the instability in the
initial opinion distribution near the boundaries. Also, $n_c$ increases
as $\et{\to}0$ in a series of bifurcations~\cite{ben2003bifurcations}.
The BC mechanism has many extensions, such as vectorial
opinions~\cite{fortunato2005vector} and coupling with a constant
external field~\cite{gonzalez2010spontaneous}.

Let us now consider the case in which agents with different opinions
have the task to form a consensual product. For this we can couple
BC with a medium considered as the common product on which the agents
should work collectively. In this case the medium has also a
convergence parameter $\ma{\in}[0,1]$ and tolerance $\ea{\in}[0,1]$,
such that the opinion $A{\in}[0,1]$ represented by the medium can be
modified by agents being dissatisfied with it. If $|x_i{-}A|{>}\ea$ we
update $A \mapsto A{+}\ma(x_i{-}A)$.  Conversely, agents tolerating
the current state of the common product ($|x_i{-}A|{\leq}\ea$) adapt
their view towards the medium, $x_i \mapsto x_i{+}\ma(A{-}x_i)$. Thus
agents can interact directly with each other and indirectly through
the medium, and a complex dynamics governed by competition of these 
local (direct) and global (indirect) interactions emerges. 
Such an interplay is also present in many other systems, 
ranging from surface chemical reactions \cite{Veser1993} and sand dunes \cite{Nishimori1993} 
to arrays of chaotic electrochemical cells \cite{Kiss2002}.

All opinions are first initialized uniformly at random and the
original BC algorithm is run until opinion groups are formed.
Then $N$ pairs of agent-agent and agent-medium interactions are
performed in each time step $t$, with agents being selected uniformly
at random.  If all agents fall within the tolerance level of the
medium the dynamics is frozen and we call such stable state {\em
consensus}. The cumulative amount of conflict or controversy in the
system is defined as the total sum of changes in the medium,
\begin{equation}
\label{eq:sumChanges}
\textstyle
S(t) = \sum_{t'=1}^t \sum_{i=1}^N | A(i) - A(i - 1) |.
\end{equation}
This quantity is analogous to the empirical controversiality measure $M$ as
it sums up the actions of dissatisfied agents~\cite{yasseri2012dynamics}.

In what follows we will analyze two versions of this model: (i) with
the agent pool fixed, and (ii) with agents being replaced by new ones at a
certain rate. For the sake of simplicity we fix $\et{=}0.2$
(leading in general to one large mainstream and two small extremist
groups) and $\mt{=}0.5$ (implying a fair compromise of opinions).


{\it Fixed agent pool.---}
For finite $N$ and if $0{<}\ea,\ma{<}1$ the system always reaches consensus.
Let $i$ be the agent with the largest opinion $x_i$, so that a discussion
with any other agent may only lower the value of $x_i$. Consider now the
event in which agent $i$ alone modifies the medium for a number of consecutive
steps. If $A{+}\ea{<}x_i$ the medium is moved towards the opinion of the
agent by a finite amount $\ma(x_i{-}A)$. Finally, after a finite number of
steps when $x_i$ falls within the tolerance level of the medium, $x_i$ will
be lowered by a finite amount larger than $\ma\ea(1{-}\ma)$. In this way we
have devised an event of finite probability where $x_i$ is decreased, which in
turn leads to a shrunken interval for the available opinion pool. Thus the
convergence to consensus is secured and the relaxation time $\tau$ can be defined,
which, however, may be astronomical for large $N$.

\begin{figure}[t]
\centering
\includegraphics[width=\columnwidth]{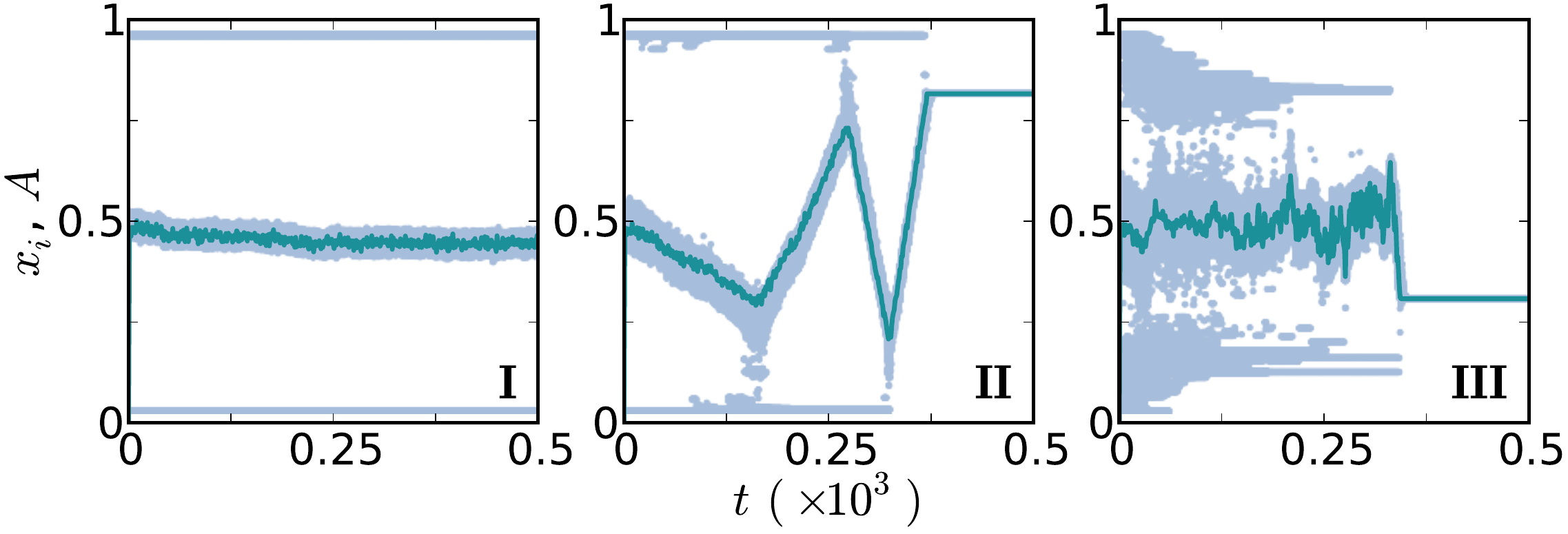}
\caption{(Color online)
Example evolutions of the agents' opinions $x_i$ (light gray/blue) and the value
$A$ of the medium (dark gray/green) for three qualitatively different regimes, corresponding to
$(\ea,\ma){=}(0.075,0.2)$ in I, $(0.075,0.45)$ in II, and $(0.15,0.7)$ in III.}
\label{fig:noiseless}
\end{figure}

If the tolerance $\ea$ is large, however, consensus may be
quickly reached in a finite number of unidirectional steps. By
decreasing $\ea$ a limiting case is reached, where the opinion of the
medium starts to oscillate between the points $1{-}\ea$ and $\ea$.
The change in the opinion of the medium should be the distance between such
points, so for a given $\ma$ the limit of oscillatory behavior is
$\eac{\equiv}1/(2{-}\ma)$. From now on we are interested in the non-trivial
case $\ea{<}\eac$.

We observed three different scenarios (see Fig.~\ref{fig:noiseless}) for
the dynamics depending on the values of $\ea$ and $\ma$: In case I the
opinion of the mainstream group fluctuates for a long time around a stable
value. This state is characterized by an astronomical relaxation time. In
case II the opinion of the mainstream group oscillates between the vicinities
of the extremists, with $\tau$ independent of $N$. In case III the extremists
converge as groups towards the mainstream opinion.

\begin{figure}[t]
\centering
\includegraphics[width=\columnwidth]{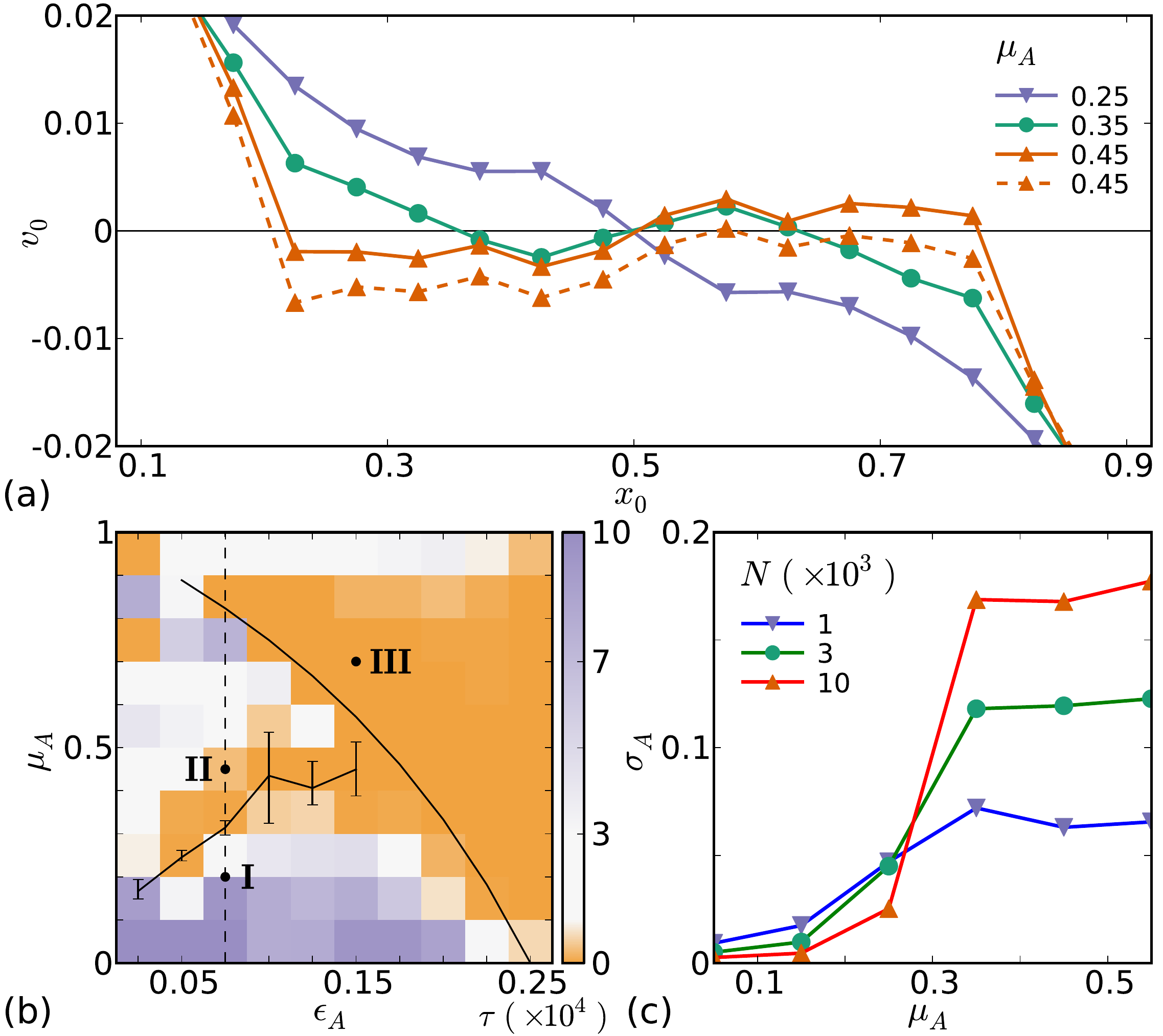}
\caption{(Color online)
(a) Velocity $v_0$ of the mainstream group at $\ea{=}0.075$ as a function of its
opinion $x_0$, for equal extremist group sizes (solid lines) and for 25\% more extremists
at $x_-$ (dashed line). (b) Phase diagram $(\ea, \ma)$ with shading indicating the
relaxation time $\tau$. Points give parameter values for the examples in Fig.~\ref{fig:noiseless}
and lines indicate boundaries between different regimes, denoted by Roman numerals. (c) Order
parameter $\sigma_A$ for the transition between I and II at $\ea{=}0.075$ (dashed line in (b)).}
\label{fig:velocity}
\end{figure}

The transition between regimes I and II can be described with stability
analysis. We use the following assumptions: First, there are three opinion
groups, one mainstream with opinion $x_0$ and two extremists with opinions
$x_-$ and $x_+$. Second, $N$ is large enough such that the change in
the opinions of the groups and correspondingly the change in the probability
distribution $\pa$ of the opinion of the medium is slow compared to a single edit.
In this case the stationary master equation for $\pa$ can be written as
\begin{equation}
\label{eq:Adiffeq}
\textstyle
0 = \sum_i \Big( -\pa(A) \Theta[ d_{A,  x_i} ] + \pa(A_i) \Theta[ d_{A_i,  x_i} ] \Big) n_i,
\end{equation}
where $n_i$ is the relative size of group $i {\in} \{0,-,+\}$, $d_{A,x_i}{=}|A{-}x_i|{-}\ea$,
and $A_i{=}(A{-}{\ma}x_i){/}(1{-}\ma)$ is the opinion of the medium
from where it would jump to $A$ after the interaction with $x_i$.
For all values of $A$ with $|x_0{-}A|{<}\ea$, the mainstream group is moved
towards $A$ with probability $n_0\pa$. The resulting velocity of
the opinion of the mainstream group is then
\begin{equation}
\label{eq:v0}
\textstyle
v_0(x_0) = V(\ma) n_0\int_{x_0 - \ea}^{x_0 + \ea} \pa(A)(A - x_0) dA,
\end{equation}
where $V(\ma)$ is a positive constant.
In Fig.~\ref{fig:velocity}(a) we show how $v_0$ depends
on $\ma$. If $n_-{=}n_+$ the opinion of the mainstream group is stable
at $x_0{=}1/2$ for low values of $\ma$, due to the negative
slope of $v_0$. Its point of stability bifurcates as $\ma$
increases and the mainstream group will drift towards one of the
extremes.  As soon as the opinions of the extremists get within the
tolerance of the medium some of them will move towards the mainstream.
When enough extremists have converted to the mainstream group the
velocity of the mainstream gets reverted (see dashed line in
Fig.~\ref{fig:velocity}(a)) and the mainstream group will head towards
the other extreme.  According to our calculations, more than 25\%
population difference between extremists is needed for the reversal.
This ensures that consensus is reached after a few oscillations, which
makes the relaxation time independent of $N$. In
Fig.~\ref{fig:velocity}(b) the numerical boundary between regimes I
and II is drawn at the marginal stability of the
mainstream group. We note here that for some values of $\et$ the shape
of the boundary between regime I and II is more complicated and may
even include islands.

After the last interaction with the extremists the opinion of the
medium and the mainstream group will remain in the vicinity of one of
the extremes. In the thermodynamic limit this leads to a symmetry
breaking in the stationary state of regime II. Conversely, in
regime I the relaxation time grows exponentially with the number
of agents, as a sequence of low probability (${\propto}1/N$) events are
needed for convergence. Thus for $N{\to}\infty$ we find a stationary
state where the opinion of the mainstream group is at 1/2. Small shifts
are possible due to differences in extremist populations, but since their
ratio determines the opinion of $A$ disturbances vanish as $1/\sqrt{N}$.
Then it is safe to define the order parameter $\sigma_A$ as the standard
deviation of the opinion of the mainstream group. When $N{\to}\infty$ this
tends to 0 for case I and increases for case II, as depicted in Fig.~\ref{fig:velocity}(c).
The latter reflects a bimodal distribution $\pa$ corresponding to the broken symmetry.

Regime III is characterized by converging extremist groups. As $\ea$ and
$\ma$ increase, the jump of the medium is big enough so that in one step
the extremists get within its tolerance interval and start drifting
inwards. Thus the step size must be $\Delta{=}\ma(1{/}2{-}\ea){=}1{/}2{-}2\ea$,
where $1/2$ is the distance between the mainstream and the extremist
groups. The boundary $\ma{=}1{-}\ea{/}(1{/}2{-}\ea)$ is shown in
Fig.~\ref{fig:velocity}(b) separating regime III from the rest.


{\it Agent replacement.---}In real systems the agent pool is often not
fixed in time as people come and go. We introduce the agent renewal
rate $\pn$ as the probability for an agent to be replaced by a new one
with random opinion before the interactions. In this section we fix
$\ma{=}0.1$ to reduce the number of parameters, focusing solely on $N$,
$\ea$ and $\pn$. Intuitively it is then clear that for $\ea{<}1/2$ and $\pn{>}0$
the dynamics never converges to a stationary state, as opposed to the
case of a fixed agent pool. This is because for any $A$ value there is
a finite probability that a new agent enters with an opinion outside the
tolerance level of the medium, after which this new agent may change the value of $A$.

In the insets of Fig.~\ref{fig:empirical} we display some examples of the
time evolution of $S$ for $N\pn{=}4$ and $\ea{=}0.47,0.46,0.44$. As in~\cite{yasseri2012dynamics}
we can distinguish three qualitatively different regimes as $\ea$ decreases:
(a) Single conflict, where $S$ is dominated by an initial increase and a
prolonged peace signaled by long plateaus. (b) A series of small
plateaus of consensus separated by conflicts. (c) A continuous
increase of $S$ indicating a permanent state of war. We define a conflict
as the period between two plateaus of $S$ and denote the number of conflicts
per unit time by $r$. The two extreme regimes are both characterized by low
values of $r$: in the peaceful regime where $\ea$ is large there are few
conflicts, while for small $\ea$ there is only one never-ending conflict.
These regimes are separated by a region full of small conflicts.

\begin{figure}[t]
\centering
\includegraphics[width=\columnwidth]{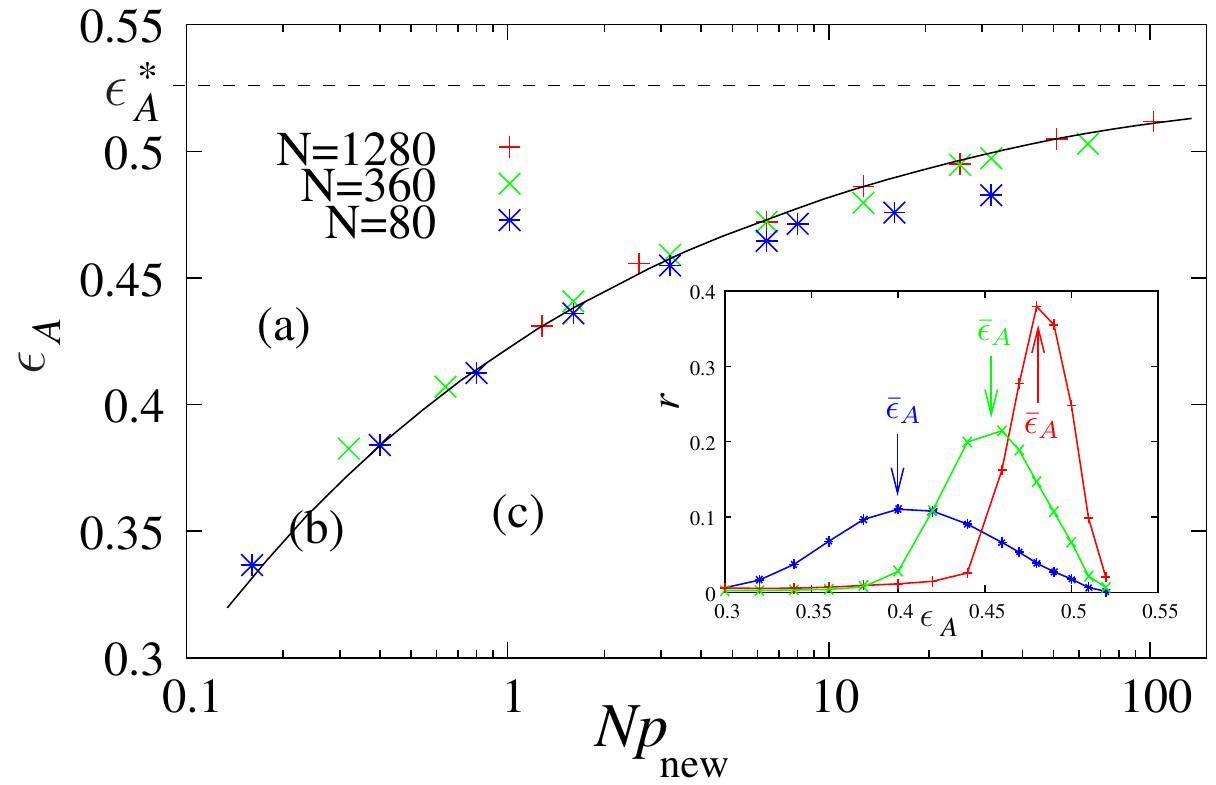}
\caption{(Color online)
Phase diagram $(N\pn, \ea)$ in the case of agent replacement. The transition
point $\eat$ shows agreement between numerical (points) and analytical (line)
results. Letters denote regimes of conflict and correspond to labels in
Fig.~\ref{fig:empirical}. Inset: Conflict rate $r$ as a function of $\ea$
for varying $N$ at $\pn{=}0.01$.}
\label{fig:noisy}
\end{figure}

In the inset of Fig.~\ref{fig:noisy} we show the variation of $r$ with $\ea$.
As $N$ increases, regimes (a) and (c) are indeed separated by a thinning transition
region (b) of many conflicts. A critical tolerance value between consensus and controversy
regimes is then identified by a maximum in the conflict density $r$ and is denoted by
$\eat$.

We note that both $\eat$ and $r(\eat)$ increase for larger
$N$, but true divergence associated with critical behavior
near a phase transition cannot be observed here due to the condition
$r{\leq}1$. The transition point $\eat$ depends both on $N$ and $\pn$,
and its corresponding phase diagram is shown in Fig.~\ref{fig:noisy}. 
The transition between peace and conflict can be derived from the
matching of two timescales: (i) the relaxation time of the system
without agent renewal, and (ii) the time scale of agent renewal.
If the latter is too small no relaxation takes place and we have
an ever-present conflict. Thus at the transition point both timescales should be equal.

The relaxation time $\tau$ for a fixed agent pool can be calculated
in regime III if $\ea{>}0.25$ (for details see Supplementary Material).
In this case $A$ can make only few jumps up and down. Knowing the
distribution of the opinion of the agents the task is to eliminate the
extremists. The rate equations for the medium and extremist movement can be
established and solved analytically to give the mean relaxation time as
\begin{equation}
\tau = cN\left([2 e^2+e_0^2  (n-1)] n - e e_0 (n-1) (2 + n)\right),
\end{equation}
where $e{=}\eac-\ea$, $e_0{=}\eac-1/2$, $c$ is a constant depending on
$\ma$, and $n$ denotes the integer part of $e/e_0$ counting the number
of steps the medium can make in one direction.

The number of new agents per unit time is $N\pn$, so we expect the
transition point $\eat$ to be at $1{=}N\pn\tau(\eat)$, shown in Fig.~\ref{fig:noisy}
as a continuous line. Such result is in considerable agreement with
the numerical computation of $\eat$, with one single fit parameter.
It also holds for other values of $\ma$, deviations appearing
only for $\ma\ll0.01$. We expect that in the $\pn{\to}0$
limit $\eat{\equiv}0$, as there is no state with permanent conflict in the
fixed agent pool case. Furthermore, for $\pn{\to}\infty$ we have $\eat{=}\eac$,
as this is the point above which no position of the medium allows for a conflict.
The curves for different system sizes fall upon each other if the number of
new agents per unit time is used as control parameter irrespective of the total
number of agents. This means that consensus is as vulnerable to many people as
it is to few.

Overall, agent replacement for the non-trivial case $\ea{<}\eac$ gives a
transition between regime (a) representing practically peace ($dS/dt{\ll}1$)
and regime (c) representing continuous conflict ($dS/dt\simeq1$). The transition
can happen if many new agents enter the system either by increasing $\pn$
or $N$. An analogue of this transition is indeed observed in Wikipedia, namely
that a peaceful article can suddenly become controversial when more people get
involved in its editing~\cite{ratkiewicz2010characterizing,yasseri2012dynamics}.


{\it Summary.---}A general question addressed by our extended opinion formation
model is the competition and feedback loop between direct agent-agent interactions
and the indirect interaction of agents with a `mean field' collectively created.
We find that convergence is always reached when the indirect interaction mechanism
is present, even in situations in which the agent-agent interaction alone does not
lead to it. We have also described different dynamical regimes of approaching
convergence, finding in particular that the transition between regimes I and II for
a critical value of the convergence parameter $\ma$ involves a symmetry-breaking
mechanism for the collectively created opinion $A$. In the case of agent replacement
we find a transition from a relatively peaceful situation to a perpetual state of
conflict when the rate of replacement is increased above a threshold. Such finding
is in agreement with different conflict scenarios observed in Wikipedia. This
first step in comparing an opinion model with real data calls to extend the model
by including networked interactions between agents, individual tolerances,
heterogeneously-distributed times between successive edits, and external events to
enable a more quantitative comparison between the model and empirical observations.

JT, TY, KK, and JK acknowledge support from EU's FP7 FET Open STREP
Project ICTeCollective No. 238597, and JK also support by FiDiPro
program of TEKES. MSM acknowledges support from project FISICOS
(FIS2007-60327) of MINECO. GI acknowledges support from an STSM
within COST Action MP0801 and is grateful to IFISC for hospitality.

\bibliography{literature}

\section{Supplementary Material}

{\it Derivation of the relaxation time.---}We calculate the relaxation
time $\tau$ for a fixed agent pool if $\ea{>}0.25$. We use a mean-field
approach and assume that in the ensemble average the distribution of
agents is homogeneous along $[0,1]$, since the relaxation time $\tau$ is
also defined as an ensemble average. We consider the initial opinion of
the medium  to be random and uniformly distributed, and without loss of
generality starting from $A{\in}[1/2,\eac]$. Finally, let use define
$e{\equiv}\eac{-}\ea$ and $e_0{\equiv}\eac{-}1/2$.

We divide $\ea{\in}[0.25,\eac]$ into small intervals within which the
medium can only make a given number of steps up and down. The first
is $\eac{>}\ea{\geq}1/2$, where the opinion of the medium is either stable or
can only make a jump up and down. In this case there is an interval
$|A-1/2|{<}\ea{-}1/2$ for the opinion of the medium at which it covers
the whole opinion range. Then consensus is immediately
reached and the contribution to $\tau$ is zero. Thus the probability
to have an oscillating state is
\begin{equation}
p_\mathrm{osc}=\frac{e}{\eac-1/2}=e/e_0.
\end{equation}

Consensus is hindered by extremists. The number of extremists
in this case is on average proportional to $e(2-\ma)N$ and is
distributed between the two extremist groups. Mainstream agents who are
always within the tolerance of the medium can be disregarded since
they never change it. If the opinion of the medium is on one
side of the middle and an extremist there is chosen for editing,
the agent will change its opinion towards the medium since it lies
within the tolerance level, thus leaving the group. Therefore
the extremist group loses one member if it is chosen repeatedly without
choosing agents from the other extremist group.

The problem can be traced back to the following urn model. Let us have
$M{=}eN{/}\eac$ balls and choose $m_1{\in}[0,M]$ randomly from a distribution
with mean around $M/2$ and standard deviation $\sigma{=}\sqrt{M}$. Then put
$m_1$ balls into one urn and the rest into another. An urn $i{=}1,2$ is chosen
with probability $m_i{/}(m_1{+}m_2)$. If we chose the same urn as in the previous
turn, we remove a ball from this urn. Then, the average time one needs to empty an
urn is $\tau_\mathrm{osc}{\simeq}2M$ for large $M$.

We have to take into account that we spend some time in choosing
already satisfied agents. The rate at which we choose
extremists varies from $e{/}\eac$ to $1{/}N$, where the latter clearly
dominates and gives another $N$ dependence to the relaxation time.
Since $\tau$ is defined in units of $N$ edits, the convergence time is
given by
\begin{equation}
\tau_0=p_\mathrm{osc}\tau_\mathrm{osc}
= ce^2N,
\end{equation}
where $c$ is a constant.

\begin{figure}[t]
\includegraphics[width=\columnwidth]{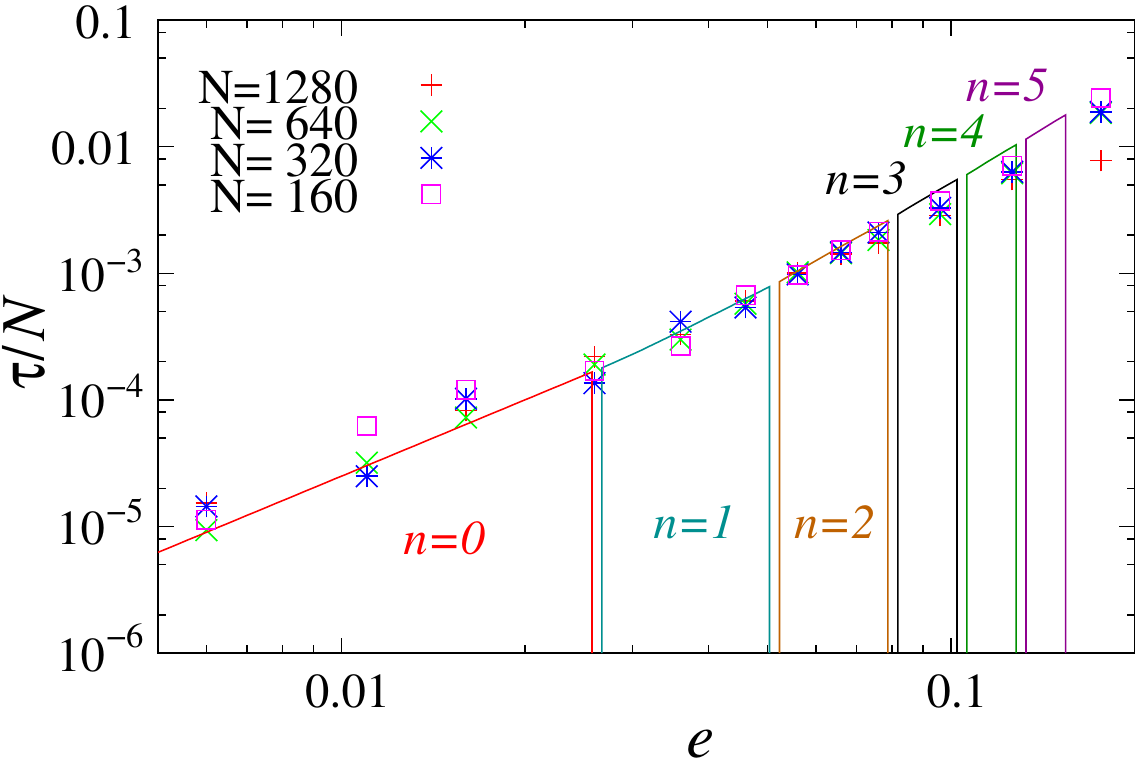}
\caption{Scaled relaxation time $\tau{/}N$ as a function of $e{=}\eac{-}\ea$,
for several values of the integer $n$ and system size $N$. Points are numerical
results and lines analytical calculations.}
\label{Fig:initcond2}
\end{figure}

Let us now consider the case $0.5{-}e_0{<}\ea{<}0.5$, where
any initial position of $A$ may lead to oscillations. If the initial
position of $A$ is such that $|1/2-A|{>}1/2-\ea$, then the previous urn model
applies giving $\tau_a{=}ceNe_0$. On the other hand,
if $|1/2-A|{<}1/2-\ea$ the medium needs two jumps from covering one
extreme to the other, yet the jump probabilities are not equal.
Jumping from the middle is much less probable since there are fewer
agents left out. So the jump from the middle will be a bottleneck
and determine the relaxation time $\tau_b$, calculated from the previous
urn model with asymmetric jump probabilities as $\tau_b{=}c(2e-e_0)N$.

This double oscillation happens with probability proportional to
$p_b{=}e{-}e_0$. When completed, the medium can only make one jump back
and forth and we return to the case of $\tau_a$. Thus the total
relaxation time is
\begin{equation}
\tau_1=\tau_a+p_b\tau_bN=c(e-e_0)(2e-e_0)N.
\end{equation}

It is straightforward to generalize this reasoning, which for
$ne_0{<}\ea{<}(n+1)e_0$ and integer $n$ gives
\begin{eqnarray}
\tau_n&=&cN(e-ne_0)[(n+1)e-ne_0]+c\sum_{i=1}^{n} [ie-(i-1)e_0]e_0\cr
&=& cN\left([2 e^2+e_0^2  (n-1)] n - e e_0 (n-1) (2 + n)\right).
\end{eqnarray}
Fig.~\ref{Fig:initcond2} shows good agreement between numerical and
analytical calculations of the relaxation time in the case of a fixed agent pool.

\end{document}